\documentclass[12pt]{article}

\newcommand{\x}{arXiv:}
\newcommand{\m}{\mathrm}

\usepackage{graphicx}

\begin{document}
\thispagestyle{empty}
\begin{center}

\null \vskip-1truecm \vskip2truecm

{\Large{\bf \textsf{A Universal Lower Bound on the Specific Temperatures of AdS-Reissner-Nordstr$\m{\ddot{o}}$m Black Holes with Flat Event Horizons}}}

{\large{\bf \textsf{}}}

{\large{\bf \textsf{}}}

\vskip1truecm

{\large \textsf{Brett McInnes
}}

\vskip1truecm

\textsf{\\  National
  University of Singapore}

\textsf{email: matmcinn@nus.edu.sg}\\

\end{center}
\vskip1truecm \centerline{\textsf{ABSTRACT}} \baselineskip=15pt
\medskip

We show that, in a gravitational theory [in any number of dimensions greater than 3] which admits BPS branes and AdS-Reissner-Nordstr$\ddot{\m{o}}$m black holes with flat event horizons, the specific [dimensionless] temperature of such a black hole is bounded below by $\approx$ 0.156875. This confirms the recent suggestion by Hartnoll and Tavanfar, to the effect that no such black hole can be arbitrarily cold, since from the AdS/CFT dual point of view the low-temperature degrees of freedom should not be concealed by the equivalent of an event horizon.

\newpage

\addtocounter{section}{1}
\section* {\large{\textsf{1. Cold Black Holes}}}
Perhaps the most important property of electrically neutral AdS static black holes is that \emph{they cannot be arbitrarily
cold.} Instead, as their temperature drops, they undergo a Hawking-Page \cite{kn:hawpa} phase transition\footnote{This is a familiar story for AdS black holes with \emph{spherical} event horizons. However, similar statements hold ---$\,$ though the details differ in interesting ways  ---$\,$ when the event horizon is flat \cite{kn:surya}.} to a quite different geometry, one with no horizon. It is this fact that permits the celebrated dual interpretation \cite{kn:confined} of these black holes in terms of a field theory which may describe aspects of the quark-gluon plasma \cite{kn:edshur} phase of quark matter [in the limit of zero chemical potential]. For it is clear that such a plasma cannot exist at low temperatures. Notice that this effect excludes not just zero-temperature black holes, but also some holes with small but positive temperatures.

In fact, Hartnoll and Tavanfar \cite{kn:harttav} have generalized this observation: they remark that a black hole description is entirely inappropriate for dual field theories at low temperatures, since the very existence of an event horizon means that the field theory degrees of freedom are in some sense inaccessible in that regime. When formulated in this way, the problem of understanding cold black holes goes far beyond the much-debated status of the third law of black hole thermodynamics [see for example \cite{kn:SWH}]. For the conclusion to be drawn from this line of reasoning is not merely that arbitrarily cold black holes with dual descriptions should always have a correspondingly small value for the entropy ---$\,$ instead, one should conclude that such black holes are always replaced, at sufficiently low but non-zero temperatures, by some other geometries without horizons [such as the ones described in \cite{kn:erik} and references therein]. The existence of the Hawking-Page transition confirms this.

Unfortunately, this satisfactory property of the uncharged case apparently fails to persist for charged static black holes: the temperature can, classically, be made arbitrarily small simply by adjusting the charge, and this can be done for any value of the black hole mass. Now charging the black hole corresponds, on the field theory side of the duality, to introducing a non-zero chemical potential. The problem on this side is still more clear: the quark-gluon plasma does not exist at low temperatures in this case either ---$\,$ in fact, it is quite possible \cite{kn:alford} that the phase transition occurs at \emph{higher} temperatures when the chemical potential is very large. If we follow the above argument, we should conclude that some non-classical effect must intervene to cause these black holes to become unstable as their temperature drops; furthermore, this should happen \emph{before} extremality [which corresponds to zero temperature] is attained. The point is that this effect must exclude not just zero-temperature black holes, but even some which are quite ``hot", in a sense which must be specified. In short, the existence of arbitrarily cold static black holes of fixed non-zero mass leads to unphysical behaviour on both sides of the duality.

In order to make sense of this situation, we need some kind of lower bound on the possible temperatures of static AdS black holes [of a given mass] as their charge is increased. We should expect this bound to depend on the mass of the black hole: very massive black holes of this kind are very hot when they are not charged, so we expect the temperature bound to be effective [as they are charged up, causing their temperatures to fall] at higher temperatures for them than for low-mass holes\footnote{We are dealing here exclusively with black holes which are static even when Hawking radiation is taken into account, that is, with the ``eternal" black holes \cite{kn:juan} that exist in the asymptotically AdS case. These black holes have a positive specific heat; their temperature \emph{increases} with their mass.}. Since the mass of the hole is related to the field-theory energy density, this means that the effect we need here is most prominent in the field theory at high densities, which seems reasonable. [In the language of the Hartnoll-Tavanfar argument: high-density, high-entropy systems, with many excited degrees of freedom, should be less effectively concealed by an event horizon ---$\,$ that is, the horizon should dissolve at higher temperatures ---$\,$ than low-density systems.] Thus what we really need is a lower bound on what might be called the ``\emph{specific temperature}" of the hole, that is, on a [dimensionless] measure of the temperature normalized by the mass. [A precise definition will be given in Section 2, below.]

Hartnoll and Tavanfar point out that, in fact, a number of effects are known or conjectured \cite{kn:denhart} which can cause zero-temperature AdS static black holes of non-zero mass, and with flat event horizons\footnote{Recall that flat event horizons are needed here if a dual theory, if any, is to propagate on a locally flat spacetime.}, to become unstable. It would greatly simplify this picture if we could show that one of these effects ensures that all zero-temperature charged AdS static black holes of non-zero mass are unstable; if we could show, in other words, that there is a \emph{universal} explanation for the absence of such black holes. [This may be related to the ``gravity as the weakest interaction" hypothesis \cite{kn:vafa}.] Confirmation of this proposal would of course clarify the situation very considerably, particularly if we could strengthen it to the assertion that there is some universal effect which ensures that \emph{all AdS-Reissner-Nordstr$\ddot{\m{o}}$m black holes become unstable when they are ``sufficiently cold"}.

In short, then, the idea is that the non-existence of very cold AdS black holes [of a given mass] is a general phenomenon, responsible in special cases for the situation revealed by duality. By this we mean in particular that holographic dual descriptions for AdS black holes are currently available only in [relatively] low dimensions. Whether such descriptions exist in higher dimensions is a question of major importance, and one which has attracted much recent attention \cite{kn:deboer1}\cite{kn:edelstein1}\cite{kn:escobedo}\cite{kn:edelstein2}\cite{kn:deboer2}\cite{kn:edelstein3}. Settling this issue will reveal a great deal, for example, about the precise nature of the relationship between holography in general and the specific case of the AdS/CFT correspondence [see for example \cite{kn:raphbo}]. It is therefore interesting to study cold AdS$_{\m{n+2}}$-Reissner-Nordstr$\ddot{\m{o}}$m black holes, with flat event horizons, for arbitrary values of n.

We have previously \cite{kn:AdSRN} investigated one possible source of instability for AdS$_{\m{5}}$-Reissner-Nordstr$\ddot{\m{o}}$m black holes with flat three-dimensional event horizons. This arises from a very general argument, based on [but perhaps not restricted to] string theory. The latter has taught us that any consistent theory of quantum gravity almost certainly entails the possibility of probing semi-classical backgrounds using extended objects ---$\,$ branes. Seiberg and Witten \cite{kn:seiberg} showed that certain kinds of background geometries can give rise to a pair-production instability for these branes. This instability does indeed arise for n = 3 charged AdS static black hole geometries of fixed mass with flat event horizons, but only when the charge is a large fraction of the extremal charge. This is exactly what we need.

Among all of the various possible ways in which black holes dual to field theories can be unstable, this one has two crucial distinctive properties. First, it arises automatically even in the simplest possible case: ordinary AdS-Reissner-Nordstr$\ddot{\m{o}}$m black holes. One does not need to introduce R-charge or scalar fields, and so on. In this sense, Seiberg-Witten instability is, next to the Hawking-Page transition, the most fundamental and ubiquitous form of instability in this situation. Secondly, this effect becomes \emph{more} important as the chemical potential of the dual theory increases, to such an extent indeed that it dominates all other kinds when the chemical potential is sufficiently large. [In \cite{kn:triple} we argued that ``large" here means the value of the chemical potential at the triple point in the field theory phase diagram. The chemical potential of the triple point in a realistic quark matter phase diagram \cite{kn:alford} is not known, but 1000 MeV is probably a reasonable order-of-magnitude estimate.] In short, the Seiberg-Witten effect is always present in the background, even if it is only decisive at high values of the charge on the black hole. It is therefore a prime candidate for the universal effect we seek.

In summary, then, there is a lower bound on the temperatures of AdS$_{\m{n+2}}$-Reissner-Nordstr$\ddot{\m{o}}$m black holes with n = 3, with fixed mass, and with flat event horizons. The bound is determined by the Seiberg-Witten process in cases where the charge is relatively large\footnote{In fact, the situation when the charge is large has a simpler holographic interpretation than when it is small. For, in the dual picture, the real QGP does \emph{not} in fact undergo a phase transition as the temperature drops when the chemical potential is smaller than a critical value ---$\,$ instead, there is a smooth crossover \cite{kn:mohanty}. For a discussion of this regime from a holographic point of view, see \cite{kn:oliver}. In the present work, as in \cite{kn:triple}, we are concerned with larger values of the chemical potential than those considered in \cite{kn:oliver}: we are at or beyond the \emph{triple} point, not the critical point.}, and we have argued that this property of the black hole corresponds, in the dual picture, to the fact that the field theory plasma has an absolute minimum possible temperature. Unfortunately, the analysis given in \cite{kn:AdSRN} does depend on having three-dimensional event horizons, that is, n = 3. [It makes use of certain properties of a cubic polynomial, and one has a cubic in that case precisely because n = 3.] Thus it is not clear that the conclusions of that work can be extended to higher [or, indeed, lower] bulk dimensions than five .

We shall see in this work that the specific temperatures of \emph{all} AdS$_{\m{n+2}}$-Reissner-Nordstr$\ddot{\m{o}}$m black holes [with n $\geq$ 2] are bounded below by the Seiberg-Witten effect. In fact, we have a stronger and more striking result: there is an absolute lower bound which is universal, in the sense that it is independent of n. This underlines the fact that there is a fundamental reason for the non-existence of very cold static black holes in AdS, and it supports the contention of Hartnoll and Tavanfar \cite{kn:harttav}, that the physics of sufficiently cold gravitating systems, and of their dual field theories if any, should always be accessible to inspection.

We begin with the details of these black holes.

\addtocounter{section}{1}
\section* {\large{\textsf{2. AdS$_{\m{n+2}}$-Reissner-Nordstr$\ddot{\m{o}}$m Black Holes with Flat Event Horizons}}}
In applications of the AdS/CFT correspondence, we are usually interested in having a locally flat [conformal] geometry at infinity. This means that we need a Reissner-Nordstr$\ddot{\m{o}}$m black hole in the bulk with a flat, rather than spherical, event horizon. Such objects do exist in AdS \cite{kn:lemmo}; in fact, there are [\emph{exact}] asymptotically AdSRN$_{\m{n+2}}$ black hole solutions with event horizons having the topology of \emph{any} n-dimensional space of constant curvature k = $\{ -1, 0, +1 \}$. [Here and henceforth, n always satisfies n $\geq$ 2.] These solutions have the form, if the asymptotic AdS curvature is $-1$/L$^2$,
\begin{eqnarray}\label{ZA}
\m{g(AdSRN^k_{n+2})} &=& -\, \m{\Bigg[{r^2\over L^2}\;+\;k\;-\;{16\pi M\over n V[X^k_n] r^{n-1}}\;+\;{8\pi Q^2\over n(n-1)\big(V[X^k_n] \big)^2 r^{2n-2}}\Bigg]dt^2\;}
\nonumber \\
& & + \m{\;{dr^2\over {r^2\over L^2}\;+\;k\;-\;{16\pi M\over n V[X^k_n] r^{n-1}}\;+\;{8\pi Q^2\over n(n-1)\big(V[X^k_n] \big)^2 r^{2n-2}}} \;+\; r^2\,d\Omega^2[X_n^k].}
\end{eqnarray}
Here $\m{d\Omega^2[X_n^k]}$ is the n-dimensional ``angular" part of the metric; it is a metric of constant curvature k = $\{ -1, 0, +1 \}$ on an n-dimensional space X$\m{^k_n}$ with [dimensionless] volume $\m{V[X^k_n]}$. [Thus, for example, for the unit two-sphere  $\m{V[S^1_2] = 4\pi}$; for the unit n-sphere $\m{V[S^1_n] = 2\pi^{(n+1)/2}/\Gamma((n+1)/2)}$, where $\Gamma$(x) is Euler's gamma function; for the Weeks manifold W$^{-1}_3$, the three-dimensional compact manifold of constant curvature $-1$ having least volume \cite{kn:gabai}\cite{kn:gabaiagain}, $\m{V[W^{-1}_3]} \approx 0.9427$; for a cubic n-torus T$\m{^0_n}$ with angular coordinates having a common periodicity 2$\pi$K, $\m{V[T^0_n] = 2^n\pi^nK^n}$; bear in mind that $\m{V[X^k_n]}$ depends on the topology of X$\m{^k_n}$ as well as k and n.] This latter quantity has to be included in the other components of the metric in order to ensure \cite{kn:peldan} that the ADM Hamiltonian is properly normalized at infinity; thus indeed M and Q are the ADM mass and charge respectively.

In the AdS/CFT application one is usually more interested in the \emph{densities} of M and Q at the event horizon of the hole, since these have a direct interpretation in the dual theory. If r$_{\m{eh}}$ is the value of the radial coordinate r at the event horizon, then these densities are given respectively by M/$\m{V[X^k_n]}$r$\m{^n_{eh}}$ and Q/$\m{V[X^k_n]}$r$\m{^n_{eh}}$. Notice that M and Q only occur in the metric in the combinations
\begin{equation}\label{DAMN}
\m{M^* \;=\;M/V[X^k_n],\;\;\;\;\;Q^*\;=\;Q/V[X_n^k]}.
\end{equation}
Therefore r$_{\m{eh}}$ is determined by M$^*$ and Q$^*$, rather than by M and Q, and so the same is true of the densities, which are given by $\rho$ = M$^*$/r$\m{^n_{eh}}$ and $\sigma$ = Q$^*$/r$\m{^n_{eh}}$. This is important because in the case of black holes with toral event horizons, one might want to take the limit in which the volume of the event horizon is taken arbitrarily large [so that the dual theory is not constrained to propagate on a compact space]. Then M and Q must likewise become large, but the ``renormalized" mass and charge, M$^*$ and Q$^*$, remain finite. Since r$_{\m{eh}}$ is entirely determined by the renormalized mass and charge, it too remains finite; and so, therefore, do the densities. Whenever we wish to use AdS/CFT to compute quantities in the field theory, then, we should express them in terms of M$^*$ and Q$^*$.

The temperatures of these black holes are readily computed by the usual technique based on requiring that the Euclidean version of the geometry be regular. For an initial orientation let us first consider the case Q = 0. Then the temperature is
\begin{equation}\label{DOLT}
\m{T \;=\;{1 \over 4\pi}\Bigg[{(n-1)k \over r_{eh}} + {(n+1)r_{eh}\over L^2}\Bigg]},
\end{equation}
where r$_{\m{eh}}$ is related to the renormalized mass by
\begin{equation}\label{DOPE}
\m{M^* \;=\;{n \over 16\pi}\Bigg[kr_{eh}^{n-1}\;+\;{r_{eh}^{n+1}\over L^2}\Bigg]}.
\end{equation}
In the familiar k = 1 case, we see that T has a crucial property: it has a minimum [given by $\m{{1\over 2\pi L}\sqrt{n^2 - 1}}$] as a function of r$_{\m{eh}}$ and therefore as a function of M$^*$: the hole cannot be arbitrarily cold. When k = 0, the case of interest in applications, the situation is somewhat more complex. The temperature now appears to be a monotonically increasing function of r$_{\m{eh}}$, which tends to zero as M$^*$ does. Nevertheless there is in fact a phase transition in this case also \cite{kn:surya}, again to a geometry with no horizons \cite{kn:gall1}\cite{kn:gall2}. Thus in both cases the temperature cannot be arbitrarily low when Q = 0.

We are interested here in finding some kind of lower bound on the temperatures of \emph{charged} AdS black holes, particularly in the k = 0 case. In order to do that, we must of course set an appropriate scale; in other words, we need a dimensionless measure of temperature. We can find the appropriate measure by examining the form taken by equation (\ref{DOLT}) when the energy density at the horizon is large.

The introduction of charge only serves to reduce the temperature given in equation (\ref{DOLT}); that is, this formula yields the \emph{maximum} possible temperature for an AdS black hole of a given renormalized mass. When M$^*$ is very large, so is r$_{\m{eh}}$, and hence so [by equation (\ref{DOPE})] is M$^*$/r$\m{_{eh}^n}$; we are in the high density regime. We then have $\m{T\,\approx \,{(n+1)r_{eh}\over L^2}}$, and then using (\ref{DOPE}) we obtain
\begin{equation}\label{DROOG}
\m{T_{max} \;\approx\;{n+1 \over 2^{n\over n+1}}\,\times\,{1\over 2^{{n-2\over n+1}}\pi^{{n\over n+1}}n^{{1\over n+1}}}\,\times \, \big[M^*/L^{2n}\big]^{{1 \over n+1}}};
\end{equation}
the reason for writing the formula in this way will be explained later. We see that the maximum temperature is a pure number multiple of the quantity $\m{\big[M^*/L^{2n}\big]^{{1 \over n+1}}}$, which of course has the same units as temperature\footnote{In the units we use here, M and M$^*$ have units of [length]$^{\m{(n-1)}}$ while T has units of [length]$^{-1}$.}. [This quantity also has a direct physical interpretation on the field theory side of the AdS/CFT duality: it is related, via equation (\ref{DOPE}), to the energy density $\rho$ = M$^*$/r$\m{^n_{eh}}$.] It is therefore natural to define a ``\emph{specific temperature}" for these black holes, given by
\begin{equation}\label{X}
\m{\tilde{T} \; =\; {T \over \big[M^*/L^{2n}\big]^{{1 \over n+1}}}.}
\end{equation}
Equation (\ref{DROOG}) then yields an expression for the maximum possible value $\m{\tilde{T}_{max}}$ of the dimensionless specific temperature; we can then say that ``cold" means ``having a specific temperature which is small compared with $\m{\tilde{T}_{max}}$".

Everything said thus far applies to all electrically neutral AdS static black holes. Henceforth we specialize to the case of AdS static black holes with flat event horizons; then all of the above approximate formulae become exact for neutral black holes, since now k = 0. The exact value of $\m{\tilde{T}_{max}}$ is then
\begin{equation}\label{DRONGO}
\m{\tilde{T}_{max} \;=\;{n+1 \over 2^{n\over n+1}}\,\times\,{1\over 2^{{n-2\over n+1}}\pi^{{n\over n+1}}n^{{1\over n+1}}}};
\end{equation}
the task now is to show that there is also a \emph{minimal} specific temperature, and to find an expression for it.

We now allow the charge to be non-zero. When the event horizon has the geometry of an n-dimensional cubic torus T$_{\m{n}}^0$, n $\geq$ 2, then the black hole metric is
\begin{eqnarray}\label{A}
\m{g(AdSRN^0_{n+2})} &=& -\, \m{\Bigg[{r^2\over L^2}\;-\;{M\over
2^{n-4}n\pi^{n-1}K^nr^{n-1}}\;+\;{Q^2\over 2^{2n-3}n(n-1)\pi^{2n-1} K^{2n} r^{2n-2}}\Bigg]dt^2\;}
\nonumber \\
& & + \m{\;{dr^2\over {r^2\over L^2}\;-\;{M\over
2^{n-4}n\pi^{n-1}K^nr^{n-1}}\;+\;{Q^2\over 2^{2n-3}n(n-1)\pi^{2n-1} K^{2n} r^{2n-2}}} \;+\; r^2\,d\Omega^2[T^0_n]}.
\end{eqnarray}
Here 2$\pi$K is the periodicity of the angular coordinates on the cubic torus, so that $\m{V[T_n^0] = 2^n\pi^n K^n}$. It is usual, though not always really necessary\footnote{In fact, field theories on compact spaces are of considerable interest: see for example \cite{kn:hands}\cite{kn:klein}.}, to take K to infinity; in that case one uses the finite quantities M$^*$ and Q$^*$, as explained above. When working on the black hole side of the duality, it is convenient to keep K finite until the end of the computation.

The entropy of such a black hole is computed in the usual way \cite{kn:lemos2}, as one quarter of the area of the event horizon; that is,
\begin{equation}\label{B}
\m{S \;=\;2^{n-2}\pi^nK^nr_{eh}^n,}
\end{equation}
where r$_{\m{eh}}$ satisfies
\begin{equation}\label{C}
\m{{r_{eh}^{2n}\over L^2}\;-\;{Mr_{eh}^{n-1}\over
2^{n-4}n\pi^{n-1}K^n}\;+\;{Q^2\over 2^{2n-3}n(n-1)\pi^{2n-1} K^{2n}} \;=\;0.}
\end{equation}
It is useful to combine these equations to obtain a relation between S and Q:
\begin{equation}\label{D}
\m{\pi Q^2L^2 \;=\;2^{4-(2/n)}(n-1) \pi^2 M L^2 K S^{(n-1)/n} \;-\; 2n(n-1)S^2.}
\end{equation}
If we fix all of the other parameters, then this relation allows us to regard Q$^2$ as a function of S; it is a decreasing function on the physical domain. The function is bounded, and so Q has a maximal value, beyond which S is not defined. This maximal value is of course the extremal value of the charge, so the latter can be found by maximising Q$^2$:
\begin{equation}\label{E}
\m{\pi Q_{ext}^2 L^2 \;=\;2^{{5n-3 \over n+1}}(n+1)n^{{1-3n \over n+1}}\Big[(n-1) \pi^2 MKL^2\Big]^{{2n \over n+1}}}.
\end{equation}
The entropy at extremality is not zero: it is
\begin{equation}\label{F}
\m{ S_{ext} \;=\;2^{{2n-2 \over n+1}}\Big[(n-1)n^{-2} \pi^2MKL^2\Big]^{{n \over n+1}}}.
\end{equation}
This is the lowest possible value of the entropy.

The temperature of the black hole is computed, as usual, by considering the Euclidean form of the black hole spacetime. The Euclidean time coordinate is compactified [so that the conformal boundary has the topology of an (n+1)-dimensional torus] with period 2$\pi$P, and the temperature is related to P. The result, after a lengthy but elementary computation, is
\begin{equation}\label{G}
\m{T\;=\;{2^{2/n}nS^{1/n}\over 4\pi^2 KL^2}\;-\;{(n-1)M\over nS}.}
\end{equation}
This is an increasing function of the entropy, if we fix all parameters but the charge. Inserting the minimal value of the entropy, from equation (\ref{F}), we find that the temperature at extremality is zero. The existence of a positive lower bound on the entropy, in the absence of any lower bound on the black hole temperature, is a well-known difficulty from the holographic point of view \cite{kn:hartnollreview}. Our objective is to remove this difficulty in as general a manner as possible.

The chemical potential of the dual theory, if any, is computed as follows. The bulk electromagnetic potential one-form, using a gauge
such that the connection is not singular [see \cite{kn:clifford}, page 416], is
\begin{equation}\label{H}
\m{A\;=\;{Q \over 2^n (n-1)\pi^n K^n}\Bigg[{1\over r^{n-1}}\;-\;{1\over r_{eh}^{n-1}}\Bigg]\,dt.}
\end{equation}
The chemical potential $\mu$ is proportional to the asymptotic value of the coefficient in this expression. The magnitude of the [negative] constant of proportionality is essentially \cite{kn:myers} the ratio of the bulk gauge coupling to the curvature scale L. It is customary to take this ratio to be of order unity\footnote{For a discussion of this ratio in the context of the actual observed quark-gluon plasma, see \cite{kn:triple}. This does not affect the main result of the present work.}. Then we have, after using relations (\ref{B}) and (\ref{D}),
\begin{equation}\label{I}
\m{\mu^2\;=\;{M\over 2^{(2n-2)/n}(n-1)\pi L^2 KS^{(n-1)/n}}\;-\;{nS^{2/n} \over 2^{(5n-4)/n}(n-1)\pi^3L^4K^2}.}
\end{equation}

In terms of the AdS/CFT correspondence, suppose that one is given a point in the field theory phase plane [plotting temperature against chemical potential, see \cite{kn:alford}]; that is, one is given specific values for T and $\mu$. Assuming that the temperature is sufficiently high that there is a black hole dual, we can regard equations (\ref{G}) and (\ref{I}) as a pair of simultaneous equations for S and M; then Q can be computed from equation (\ref{D}), and so the black hole corresponding to the field theory state can be uniquely identified.

We now fix M and think of S as a parameter; as it varies when the charge on the black hole changes, it determines a curve in the field theory phase diagram  given in parametric form by equations (\ref{G}) and (\ref{I}).
As the charge varies [as it will if, for example, one throws charged particles into the hole], the curve bends down and to the right, to lower values of the temperature and higher values of the chemical potential. If nothing intervenes to prevent it, such a curve will intersect the T = 0 axis. This happens at a value of the chemical potential corresponding to the extremal charge: it is given by
\begin{equation}\label{J}
\m{\mu_{ext} \; =\; \sqrt{{(n+1)n^{{n-3\over n+1}}\Big[(n-1)\pi^2MKL^2\Big]^{{2\over n+1}} \over 2^{{5n-3 \over n+1}}(n-1)^2\pi^3K^2L^4} }}
\end{equation}
This is in fact an \emph{upper bound} on the possible values of the chemical potential, for static black holes of given mass. As in the case of the temperature, it is useful, and more meaningful, to formulate such bounds in terms of dimensionless quantities. The relevant dimensionless combination here is the value of the chemical potential [which has the same units as the temperature] normalized by the black hole's mass: denoting this by $\tilde{\mu}$, we have
\begin{equation}\label{K}
\m{\tilde{\mu} \; =\; {\mu \over \big[M^*/L^{2n}\big]^{{1 \over n+1}}}. }
\end{equation}
The bound on the chemical potential can then be expressed in a dimensionless way as
\begin{equation}\label{L}
\m{\tilde{\mu}\;\leq\;\tilde{\mu}_{ext} \; =\; \sqrt{{(n+1)n^{{n-3\over n+1}} \over 2^{{3n-3 \over n+1}}(n-1)^{{2n \over n+1}}\pi^{{n-1 \over n+1}}}}}.
\end{equation}

\addtocounter{section}{1}
\section* {\large{\textsf{3. Brane Probes of the Black Hole Geometry}}}
In a quantum theory of gravity we should be able to probe the spacetime geometry using extended objects: branes.
Seiberg and Witten \cite{kn:seiberg}, investigating this question, generalized the results of \cite{kn:frag}, and drew attention to the possible consequences of \emph{brane pair-production}. They found that this effect, which of course does not exist in any theory based exclusively on point particles, can reveal a new form of instability in systems which otherwise appear to be stable. Importantly for us, their results are established for an \emph{arbitrary} number of spacetime dimensions; furthermore, they did not confine themselves purely to Einstein spacetimes ---$\,$ that is, they allowed for non-trivial matter configurations deforming the spacetime geometry. This latter aspect was worked out in more detail subsequently by Witten and Yau \cite{kn:wittenyau} and, in the specific case of interest to us here [with gauge fields in the bulk], by Maldacena and Maoz \cite{kn:maoz}; see also \cite{kn:porrati}. In all of these works, the action of a BPS n-brane of tension $\Theta$ wrapping a constant-radius surface in a Euclideanized spacetime is computed as follows. The brane action is taken to be proportional to the difference between the area and the [suitably normalized] volume of the Euclidean brane; for example, in the case of the Euclidean geometry corresponding to the spacetime with metric given by equation (\ref{A}) above, we have
\begin{eqnarray}\label{M}
\m{\$(r)} &=& \m{2^{n+1}\pi^{n+1} \Theta P L K^n \Bigg\{r^n\Bigg[{r^2\over L^2}\;-\;{M\over
2^{n-4}n\pi^{n-1}K^nr^{n-1}}\;+\;{Q^2\over 2^{2n-3}n(n-1)\pi^{2n-1} K^{2n} r^{2n-2}}\Bigg]^{1/2}\;}
\nonumber \\
& & \;\;\;\;\;\;\;\;\;\;\;\;\;\;\;\;\;\;\;\;\;\;\;\;\;-\; \m{\m{{r^{n+1}\,-\,r_{eh}^{n+1}\over L}\Bigg\}}.}
\end{eqnarray}

The reader may object at this point that this way of computing the action may not be complete, particularly in the case in which a gauge field is present in the bulk [as it is in the work of Maldacena and Maoz, and as it is here]. For gauge fields in this context typically arise from the dimensional reduction of some higher-dimensional metric, as in \cite{kn:chamblin}, and one might therefore expect a contribution to the brane action coming from a coupling of the brane to the gauge potential. We believe that this is a valid objection in general, but not in the specific case under consideration here. For it turns out that the crucial property of the action we need here derives from a subtle aspect of the black hole geometry very \emph{far} from the hole, where [by the no-hair principle] the electric charge [or angular momentum in the higher-dimensional spacetime] and the mass are the only relevant parameters. To explain this in detail, we proceed with the analysis of the action as given in equation (\ref{M}); we will return to this discussion after that.

The action vanishes at r = r$_{\m{eh}}$, and it is not hard to show that it is positive for values of r somewhat larger than r$_{\m{eh}}$. At large values of r, however, it is clear that there is a close competition between the contributions of the area and the volume. In fact, in that limit the leading terms from each contribution are both proportional to r$^{\m{n+1}}$, and so the question as to which dominates is decided by higher order terms. Thus the sign of the action far from the hole is settled by subtle details of the spacetime geometry \emph{in the asymptotic  region}.

In order to proceed, we can manipulate the brane action to a more useful form:
\begin{equation}\label{N}
\m{\$(r) \;=\;2^{n+1}\pi^{n+1} \Theta P L^2 K^n\Bigg\{{{Q^2r^{1-n} \over 2^{2n-3}n(n-1)\pi^{2n-1} K^{2n}}\;-\;{M \over 2^{n-4}\pi^{n-1} K^n}
\over
1\;+\;\Bigg[\,1\;-\;{ML^2\over
2^{n-4}n\pi^{n-1} K^nr^{n+1}}\;+\;{Q^2L^2\over 2^{2n-3}n(n-1)\pi^{2n-1} K^{2n} r^{2n}}\,\Bigg]^{1/2}}\;+\;{r_{eh}^{n+1}\over L^2}\Bigg\}}.
\end{equation}
We can now compute a quantity which does not depend on the detailed properties of particular branes, namely the asymptotic value of the brane action per unit brane tension:
\begin{equation}\label{O}
\m{\$(\infty)/\Theta \;=\;2^{n+1}\pi^{n+1} P L^2 K^n\Bigg\{\;-\;{M \over 2^{n-3}n\pi^{n-1} K^n}
\;+\;{r_{eh}^{n+1}\over L^2}\Bigg\}}.
\end{equation}
The results of Seiberg and Witten imply that if we had performed this calculation for an AdS black hole with a \emph{spherical} event horizon, then the result would necessarily have been positive; equally, it must be negative if the event horizon is negatively curved\footnote{For a recent interesting discussion of that case, see \cite{kn:barbon}; see also \cite{kn:fragile}.}. In the case at hand, where the event horizon is \emph{flat}, there is no general rule of this kind: the asymptotic action can be either positive or negative. It is known \cite{kn:conspiracy} that the action for branes in the spacetime of a zero-charge static black hole with a flat event horizon is positive everywhere, but it is not clear that this is still true in the charged case. If in fact the action becomes negative over an infinite range of r values, then branes can draw upon an infinite reservoir of negative free energy, and the system becomes unstable to uncontrolled brane pair-production.

To investigate this, we take the uncharged black hole and imagine increasing the magnitude of the charge without changing the mass. As we know, this decreases the entropy of the black hole, and therefore it causes r$_{\m{eh}}$ to decrease. From equation (\ref{O}) we see that this causes $\$(\infty)/\Theta$ to decrease. The question of interest here is this: can $\$(\infty)/\Theta$ reach zero before the charge reaches the extremal value? If it can, then any further increase in the charge will cause $\$(\infty)/\Theta$ to become negative, and so it will follow that the system is unstable when the charge lies between that critical value and the extremal value.

Clearly $\$(\infty)/\Theta$ can be zero if and only if it is possible for r$_{\m{eh}}$ to satisfy
\begin{equation}\label{P}
\m{r_{eh} \;=\;\Bigg[{ML^2 \over 2^{n-3}n\pi^{n-1} K^n}\Bigg]^{{1\over n+1}}}.
\end{equation}
We can compute the charge at which this happens by using equation (\ref{C}) above; the result is
\begin{equation}\label{Q}
\m{\pi Q_{max}^2 L^2 \;=\;2^{{5n-3 \over n+1}}(n-1)n^{{1-n \over n+1}}\Big[\pi^2 MKL^2\Big]^{{2n \over n+1}}}.
\end{equation}
Using equation (\ref{E}) we can compare this with the extremal charge:
\begin{equation}\label{R}
\m{Q_{max}^2/Q_{ext}^2 \;=\;{n-1 \over n+1} \Bigg[{n\over n-1}\Bigg]^{{2n\over n+1}}}.
\end{equation}
It is straightforward\footnote{Let f(x) = $\m{(1-x)^{{x-1 \over x+1}}/(1+x)}$, where x, ranging between 0 and 1/2, is a continuous proxy for 1/n. Then $\m{(x+1)ln(f(x)) = (x-1)ln(1-x) - (x+1)ln(x+1)}$ and this expression is necessarily negative because the function xln(x) is concave up for all x. Hence f(x), and therefore the expression in (\ref{R}), is always less than unity.} to show that this expression is always smaller than unity. That is, as the charge is increased, Seiberg-Witten instability occurs in the spacetime around an AdS black hole with a flat event horizon, and this happens, in every dimension, \emph{before extremality is reached}. Thus, as the notation suggests, $\m{|Q_{max}|}$ is indeed the maximal possible magnitude of the charge on a stable black hole of this kind. In AdS$_4$, the case of interest in condensed matter applications \cite{kn:harttav}, $\m{|Q_{max}|}$ is about 0.916 $\times$ $\m{|Q_{ext}|}$; in AdS$_5$, the case of interest when the field theory is defined on a four-dimensional space, it is 0.958 $\times$ $\m{|Q_{ext}|}$; and in AdS$_7$, the case studied in \cite{kn:deboer1}, it is 0.983 $\times$ $\m{|Q_{ext}|}$; we are below extremality, though not very far below it.
 
The instability has the following striking property: it arises from the behaviour of branes which are \emph{very far from the black hole}, not near to it. This is surprising at first sight, since one might have thought that the relatively large charge on the hole would have its strongest effect on the branes in the region where the electric field is strong, not weak. What happens instead is that [if the charge is larger than $\m{|Q_{max}|}$] the brane action is always \emph{positive} near to the event horizon; it then decreases slowly and can become negative only when r is large. The point is illustrated in Figure 1, where we have graphed [a positive multiple of] the action in the case of an asymptotically AdS$_5$ black hole, with the charge equal to 98.8$\%$ of the extremal value, other parameters being chosen for convenience.
\begin{figure}[!h]
\centering
\includegraphics[width=1.2\textwidth]{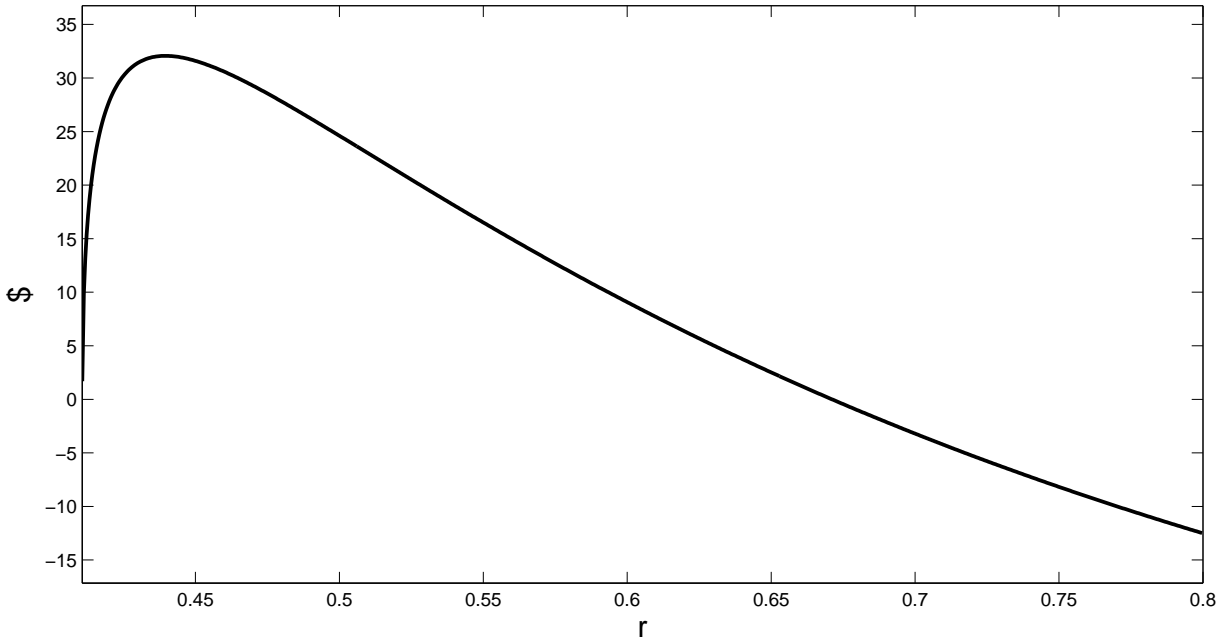}
\caption{[Scaled] Action for a Brane in AdS$_5$ Black Hole Spacetime, Q = 98.8$\%$$\m{|Q_{ext}|}$}
\end{figure}
The distance from the event horizon to the point where the graph cuts the horizontal axis depends on the extent to which $\m{|Q|}$ exceeds $\m{|Q_{max}|}$: it is large when the charge difference is \emph{small}. To put it another way: if we imagine slowly increasing the charge from a value below $\m{|Q_{max}|}$ to a value slightly larger than it, the region of negative action develops \emph{at infinity} and slowly moves inwards. [It never comes close to the event horizon, as long as the charge remains sub-extremal.]

The key point here is that the branes in the asymptotic region propagate in a geometry in which areas and [normalized] volumes differ by very small quantities, which in turn are sensitive to the asymptotic physics of the black hole.
In fact, this is a generic feature of Seiberg-Witten instability: it arises from the behaviour of branes in the asymptotic region. In this sense, it is a ``holographic" effect, and, as Seiberg and Witten showed explicitly, it is indeed always reflected in the structure of the AdS/CFT dual field theory. This has the following consequence: in the specific case of a black hole bulk, \emph{the onset of Seiberg-Witten instability can only depend on the ADM parameters}, in this case the mass and the charge [or, equivalently, the angular momentum in the compact directions if the black hole geometry is obtained by dimensional reduction]. This is in agreement with the black hole no-hair principle\footnote{The no-hair principle can in fact be violated in some cases, but one certainly expects it to hold in the present circumstances. For the current status of the no-hair theorems, see \cite{kn:bhatlah} and references therein.}.

These general comments are reflected in the structure of our computation above. Note in particular that \emph{we have not actually used the detailed form of the action function} [equation (\ref{N})] at all values of r: we only needed to use the asymptotic value [equation (\ref{O})]. This asymptotic value is determined [leaving aside the geometric parameters] by the values of P, M, and r$_{\m{eh}}$. Here P is fixed by the temperature of the black hole, which of course is determined by the ADM parameters M and Q; and the same is true of r$_{\m{eh}}$. Thus the key quantity here is fixed by M and Q, as expected.

This is relevant to our earlier discussion of the fact that there are couplings between the branes and those components of the higher-dimensional metric that generate the gauge potential. These couplings may well strongly modify the behaviour of the brane action for values of r corresponding to the region near to the black hole. However, for the reasons we have just discussed, we do not expect them to affect the asymptotic value of the action, which is all we need for the purposes of this work.

Our revision of the possible maximal charge on the black hole clearly entails an upward revision of the lowest possible entropy and temperature. For the entropy, we find, from equation (\ref{B}),
\begin{equation}\label{S}
\m{ S_{min} \;=\;2^{{2n-2 \over n+1}}\Big[\pi^2n^{-1}MKL^2\Big]^{{n \over n+1}}}.
\end{equation}
[It is easily verified that this is indeed larger than S$_{\m{ext}}$, given in equation (\ref{F}).] This is the minimum entropy in the sense that black holes with lower entropy are not stable.

Substituting this value into equation (\ref{G}) one finds at length that the minimal temperature is
\begin{equation}\label{T}
\m{ T_{min} \;=\;2^{{2-2n \over n+1}}(M/n)^{{1 \over n+1}}\Big[\pi^2KL^2\Big]^{{-n \over n+1}}}.
\end{equation}
We saw earlier that, in the absence of the Seiberg-Witten effect, there is a lower bound on the entropy but not on the temperature. This problem is now resolved: neither can be zero for any of these black holes unless the mass vanishes. An interesting way to express this fact is to multiply equations (\ref{S}) and (\ref{T}) together: the result is simply
\begin{equation}\label{U}
\m{ S_{min}T_{min} \;=\;M/n}.
\end{equation}

The existence of a non-zero lower bound on the black hole temperature entails a downward revision of the upper bound [equation (\ref{J})] on the chemical potential of the dual system. Substituting S$_{\m{min}}$ from (\ref{S}) into equation (\ref{I}), one finds after a long computation that
\begin{equation}\label{V}
\m{\mu_{max} \;=\;\sqrt{{M^{{2\over n+1}}n^{{n-1\over n+1}}\over 2^{{5n-3 \over n+1}}(n-1)\pi^{{3n-1\over n+1}}[KL^2]^{{2n\over n+1}}}}}.
\end{equation}
It is not difficult to verify that this is indeed smaller than the value given in equation (\ref{J}).

The ratio of T$_{\m{min}}$ to $\mu_{\m{max}}$ is a pure number,
\begin{equation}\label{W}
\m{T_{min}/\mu_{max} \;=\;\sqrt{{2(n-1)\over \pi n}}}.
\end{equation}
In the dual field theory, this means that the instability sets in as a system crosses a diagonal straight line in the phase diagram, a line with slope given by the right hand side of this relation. Notice that this means that even very hot systems can be unstable in this way, provided that the chemical potential is sufficiently large. This unusual property of Seiberg-Witten instability, as it arises in the case of these black holes, prompted us in \cite{kn:triple} to propose that the line given by equation (\ref{W}) might, when n = 3, be identified with the line in the quark matter phase diagram [see \cite{kn:alford}] which extends upward and to the right from the quark matter triple point.

We can now express our findings in terms of the dimensionless specific temperatures and chemical potentials: we have
\begin{equation}\label{Y}
\m{\tilde{T}\;\geq \;\tilde{T}_{min}\;=\;{1\over 2^{{n-2\over n+1}}\pi^{{n\over n+1}}n^{{1\over n+1}}}}
\end{equation}
and
\begin{equation}\label{Z}
\m{\tilde{\mu}\;\leq \;\tilde{\mu}_{max}\;=\;\sqrt{{n^{{n-1\over n+1}}\over 2^{{3n-3\over n+1}}(n-1)\pi^{{n-1\over n+1}}}}}.
\end{equation}
Comparing equations (\ref{DRONGO}) and (\ref{Y}), we see that the maximal and minimal specific temperatures differ by a factor of $\m{(n+1)/2^{n\over n+1}}$.

If we need the minimal temperature itself, we can obtain it from $\m{\tilde{T}_{min}}$ if we can express the quantity $\m{M^*/L^{2n}}$ [which occurs in the definition of $\m{\tilde{T}}$, see equation (\ref{X})] in terms of objects defined entirely on the field theory side of the correspondence. This can be done as follows. As explained above, if the plasma is cooled to a sufficient degree it must, at some special density $\bar{\rho}$ [which depends of course on its initial temperature and chemical potential] be replaced by some other state. According to our proposal, this happens when the radial coordinate at the event horizon of the dual black hole satisfies equation (\ref{P}). Using this equation and the relation $\rho$ = M$^*$/r$\m{^n_{eh}}$, we can compute the value of $\m{M^*/L^{2n}}$ when $\rho$ =  $\bar{\rho}$: it is given by
\begin{equation}\label{ZOO}
\m{{M^* \over L^{2n}}\;=\;\bigg[{8\pi \over n}\bigg]^n \,[\bar{\rho}]^{n+1}}.
\end{equation}
In principle, then, $\m{M^*/L^{2n}}$ is fixed by measuring the field-theoretic quantity $\bar{\rho}$; the actual temperature of the black hole [which is on the brink of becoming unstable], and therefore of the plasma, can then be computed. To put it slightly differently: the theory predicts a definite relationship between the temperature and the density at points in the field theory phase diagram along the line where the plasma ceases to exist.

Returning to equation (\ref{Y}): we have succeeded in showing that the temperatures of static AdS black holes, of given mass, with flat event horizons cannot be made arbitrarily small; there must be some other description of the dual system at low temperatures, one that does not involve horizons. This is true in \emph{all} spacetime dimensions $\geq$ 4.

In fact, we can strengthen this statement in an interesting way by noting the following. The expression on the right side of equation (\ref{Y}), regarded as a function of n, steadily decreases from n = 2 until a large value of n [n = 69 is the integer giving the smallest value]. After that, it \emph{increases} towards its asymptotic value, 1/2$\pi$. Thus $\m{\tilde{T}}$ cannot be arbitrarily small, in any number of dimensions. A simple numerical investigation shows that, for all integral values of n $\geq$ 2,
\begin{equation}\label{BETA}
\m{\tilde{T}\;\geq \;\approx 0.156875}.
\end{equation}
That is, \emph{the specific temperature of every AdS$_{\m{n+2}}$ static black hole, n $\geq$ 2, with a flat event horizon, is bounded below by this universal constant}\footnote{The function on the right side of (\ref{Y}) actually decreases from its value at n = 2 towards this constant very rapidly at first; even at n = 5 [that is to say, for black holes in AdS$_7$], it has already declined to approximately 0.208304.}.

For completeness let us note in closing that the dimensionless chemical potential is bounded above by the maximal value of the right side of equation ({\ref{Z}), which occurs when n = 2; therefore, for all n $\geq$ 2 we have
\begin{equation}\label{ALPHA}
\m{\tilde{\mu}\;\leq \;{1\over 2^{1/3}\pi^{1/6}}\;\approx \; 0.6558}.
\end{equation}

\addtocounter{section}{1}
\section* {\large{\textsf{4. Conclusion}}}
The study of black holes in high dimensions has recently attracted a great deal of attention in its own right; both analytical \cite{kn:harvey} and numerical \cite{kn:witek} investigations have revealed much that is surprising and interesting. AdS-Reissner-Nordstr$\m{\ddot{o}}$m black holes with flat event horizons are of particular interest in ``applied holography", and so it is important to investigate their properties, too, in arbitrary spacetime dimensions.

Classically, these black holes, like their counterparts with spherical event horizons, can have arbitrarily low specific temperatures; but we have found here that, when semi-classical effects connected with branes are taken into account, there is a universal lower bound on the specific temperatures of these objects, given approximately by 0.156875. This universal effect may be responsible for the fact, observed by Hartnoll and Tavanfar \cite{kn:harttav}, that AdS black holes never give a good description of thermal field theory configurations at temperatures which are sufficiently low relative to the energy density.

While Seiberg-Witten instability thus provides a universal account of this situation, in practice it has to compete with other, better-known sources of instability in this system. For example, these black holes are subject to a Hawking-Page transition, which sets in at a low but \emph{constant} temperature \cite{kn:surya} ---$\,$ that is, it occurs along a horizontal straight line in the phase diagram. Since the instability studied here occurs along a diagonal line in that diagram [given by equation (\ref{W})], we see that the Hawking-Page transition dominates when the chemical potential is small. In fact, in this regime one should consider the effects of R-charge on black holes \cite{kn:yamada1}\cite{kn:yamada2} and, especially, of scalar fields, which give a still better account of the region around the critical point \cite{kn:oliver}; again, these effects will dominate over the Seiberg-Witten effect in this region of the diagram. In the same way, however, the reverse is true at higher values of the chemical potential: the Seiberg-Witten effect will probably destabilize the black hole at such high temperatures that neither the Hawking-Page transition nor the scalar field effect has an opportunity to operate. Our conjecture is that this ``switch" happens at the quark matter triple point \cite{kn:triple} [or at the triple point with the largest value of the chemical potential, should there be more than one].

We remind the reader that observational probes of these extreme conditions are not entirely out of the question: there is reason to hope that this region of the quark matter phase diagram can be explored either directly \cite{kn:hohne} or perhaps by means of observations of neutron stars \cite{kn:alford2}, or possibly even by cosmological observations \cite{kn:tillmann}. However, in this work, the point we wish to stress is the extreme generality of the main result. We have not needed to postulate the presence of R-charges or of some scalar field with an exotic potential: we have worked with the simplest possible gravitational system that can have a dual interpretation in terms of a thermal field theory with a non-zero chemical potential ---$\,$ that is, with ordinary AdS-Reissner-Nordstr$\m{\ddot{o}}$m black holes. Our objective here has been to explore whether the good behaviour of this system, when branes are taken into account, extends to higher dimensions. We have found that it does, and that in fact there is actually a whole range of completely forbidden values [inequality (\ref{BETA})] for the specific temperatures of these objects. It would be interesting to investigate the consequences for recent work on duality in the higher-dimensional case
\cite{kn:deboer1}\cite{kn:edelstein1}\cite{kn:escobedo}\cite{kn:edelstein2}\cite{kn:deboer2}\cite{kn:edelstein3}.
This will require a more detailed study of the thermodynamics of high-dimensional black holes with flat event horizons; for some recent work relevant to this question, see \cite{kn:zou}.

\addtocounter{section}{1}
\section*{\large{\textsf{Acknowledgement}}}
The author wishes to thank Prof. Soon Wanmei for helpful discussions.

\end{document}